\documentclass[]{aastex631}
\usepackage{xcolor}
\usepackage{amsmath}
\shorttitle{Blackhole mass and broad emission line}
\shortauthors{Xiao et al.}
\graphicspath{{./}{figures/}}
\begin{document}

\title{The relativistic jet and its central engine of \textit{Fermi} blazars}

\correspondingauthor{Junhui Fan}
\email{fjh@gzhu.edu.cn}

\author[0000-0001-8244-1229]{Hubing Xiao}
\affiliation{Shanghai Key Lab for Astrophysics, Shanghai Normal University \\
Shanghai, 200234, China}

\author{Zhihao Ouyang}
\affiliation{School of Physics and Materials Science, Guangzhou University \\
Guangzhou, 510006, China}

\author{Lixia Zhang}
\affiliation{Center for Astrophysics, Guangzhou University \\
Guangzhou, 510006, China}
\affiliation{Key Laboratory for Astronomical Observation and Technology of Guangzhou \\
Guangzhou 510006, China}
\affiliation{Astronomy Science and Technology Research Laboratory of Department of Education of Guangdong Province \\
Guangzhou, 510006, China}

\author{Liping Fu}
\affiliation{Shanghai Key Lab for Astrophysics, Shanghai Normal University \\
Shanghai, 200234, China}

\author{Shaohua Zhang}
\affiliation{Shanghai Key Lab for Astrophysics, Shanghai Normal University \\
Shanghai, 200234, China}

\author{Xiangtao Zeng}
\affiliation{Center for Astrophysics, Guangzhou University \\
Guangzhou, 510006, China}
\affiliation{Key Laboratory for Astronomical Observation and Technology of Guangzhou \\
Guangzhou 510006, China}
\affiliation{Astronomy Science and Technology Research Laboratory of Department of Education of Guangdong Province \\
Guangzhou, 510006, China}

\author{Junhui Fan}
\affiliation{Center for Astrophysics, Guangzhou University \\
Guangzhou, 510006, China}
\affiliation{Key Laboratory for Astronomical Observation and Technology of Guangzhou \\
Guangzhou 510006, China}
\affiliation{Astronomy Science and Technology Research Laboratory of Department of Education of Guangdong Province \\
Guangzhou, 510006, China}



\begin{abstract}
Jet origination is one of the most important questions of AGN, yet it stays obscure.
In this work, we made use of information of emission lines, spectral energy distributions (SEDs), \textit{Fermi}-LAT $\gamma$-ray emission, construct a blazar sample that contains 667 sources.
We notice that jet power originations are different for BL Lacs and for FSRQs.
The correlation between jet power $P_{\rm jet}$ and the normalized disk luminosity $L_{\rm Disk}/L_{\rm Edd}$ shows a slope of -1.77 for BL Lacs and a slope of 1.16 for FSRQs.
The results seems to suggest that BL Lac jets are powered by extracting blackhole rotation energy, while FSRQ jets are mostly powered by accretion disks.
Meanwhile, we find the accretion ratio $\dot{M} / \dot{M}_{\rm Edd}$ increase with the normalized $\gamma$-ray luminosity.
Base on this, we propose a dividing line, ${\rm log} (L_{\rm BLR}/L_{\rm Edd}) = 0.25 \ {\rm log} (L_{\rm \gamma}/L_{\rm Edd}) - 2.23$, to separate FSRQs and BL Lacs in the diagram of $L_{\rm BLR}/L_{\rm Edd}$ against $L_{\rm \gamma}/L_{\rm Edd}$ through using the machine learning method, the method gives an accuracy of 84.5\%.

In addition, we propose an empirical formula, $M_{\rm BH}/M_{\rm \sun} \simeq L_{\rm \gamma}^{0.65}/21.46$, to estimate blackhole mass based on a strong correlation between $\gamma$-ray luminosity and blackhole mass.
Strong $\gamma$-ray emission is typical in blazars, and the emission is always boosted by a Doppler beaming effect.
In this work, we generate a new method to estimate a lower-limit of Doppler factor $\delta$ and give $\delta_{\rm BL Lac} = 7.94$ and  $\delta_{\rm FSRQ} = 11.55$.
\end{abstract}



\section{Introduction} \label{sec:intro}

Active galactic nuclei (AGNs) are the most energetic and persistent extragalactic objects in the universe. 
Blazars exhibit extreme observation properties, including rapid and high amplitude variability, high and variable polarization, strong and variable $\gamma$-ray emissions, or apparent superluminal motions, etc \citep{Wills1992, Urry1995, Fan2002, Fan2004, Kellermann2004, Rani2013, Fan2014, Lyutikov2017, Xiao2019}. 
The extreme observational properties result from Doppler beaming effect caused by relativistic jets \citep{Xiao2015, Pei2016, Fan2017}.

Blazars, typically, are hosted by elliptically galaxies and powered by the central supermassive black holes \citep{Urry2000, Shaw2012}.
The broadband spectral energy distribution (SED) of blazar forms a two-hump structure, which the lower energy bump is explained by the synchrotron mechanism and the higher energy bump is attributed to the inverse Compton (IC) scattering in a leptonic scenario.  

There are two subclasses of blazars that are characterized based on the emission line strength of optical spectra, namely BL Lacertae objects (BL Lacs) and flat-spectrum radio quasars (FSRQs).
The former one characterizes a spectrum with no or weak emission lines (rest-frame equivalent width, ${\rm EW < 5 \AA}$), while the latter one shows strong emission line features of ${\rm EW \ge 5 \AA}$ \citep{Urry1995, Scarpa1997}.
However, an arbitrary classification base on EW is inadequate. 
On one hand, a Doppler boosted non-thermal continuum could swamp out spectral emission lines \citep{Blandford1978, Xiong2014}. 
On the other hand, EW greater than 5 ${\rm \AA}$ may be the result of a particular low-state of jet activity.
Other indicators have been proposed to divide blazars into subclasses.
\citet{Ghisellini2011} and \citet{Sbarrato2012} suggested a distinction of accretion ratio based on the luminosity of broad-line region (BLR) measured in Eddington units, $L_{\rm BLR}/L_{\rm Edd} \sim 10^{-3}$ or $L_{\rm BLR}/L_{\rm Edd} \sim 5 \times 10^{-4}$, to separate FSRQs and BL Lacs.
\citet{Abdo2010} and \citet{Fan2016} used synchrotron peak frequency ($\rm log \nu_{s}$) to divide blazars into low-synchrotron-peaked blazars (LSP), intermediate-synchrotron-peaked blazars (ISP) and high-synchrotron-peaked blazars (HSP) and got compatible results of separating boundaries.

Emission lines with half-maximum-full-width (FWHM) greater than 1000 km/s are called broad emission lines, otherwise, narrow emission lines.
The broad emission lines are employed to estimate the central blackhole (BH) mass ($M_{\rm BH}$) by using BLR distance and FWHM assuming the BLR clouds being gravitationally bound by the central BH.
The BLR distance can be interpreted through an empirical relation between BLR distance and ionizing luminosity or through reverberation mapping \citep{Wandel1999, Kaspi2000, Kaspi2005}.
The reverberation mapping method, requires continuous observations on both continuum and emission line variations, gives a more accurate BLR distance than distance-luminosity correlation.
\citet{Kaspi2000} calibrated empirical distance-luminosity correlation by using a reverberation-mapped sample and got $R_{\rm BLR} \propto L^{0.7}_{5100}$, here $L_{5100}$ is the continuum luminosity at $\lambda = 5100 \ {\rm \AA}$. 
\citet{Greene2005} noticed that the emission line luminosities of H$\alpha$ and H$\beta$ have a strong correlation with $L_{5100}$.
They substituted the $L_{5100}$ with $L_{\rm H \alpha}$ and $L_{\rm H \beta}$, and suggested $M_{\rm BH} \propto \ L_{\rm H \alpha}^{0.55}$ and $M_{\rm BH} \propto \ L_{\rm H \beta}^{0.56}$.
$\rm Mg_{II}$, $\rm C_{IV}$ were also explored by other researchers \citep{McLure2004, Vestergaard2006, Vestergaard2009, Shen2011, Shaw2012}.
For some sources without broad emission lines, especially BL Lac objects, their $M_{\rm BH}$ can be estimated from the properties of their host galaxies with $M_{\rm BH}-\sigma_{\star}$ and $M_{\rm BH}-L$, where $\sigma_{\star}$ and $L$ are the stellar velocity dispersion and the bulge luminosity \citep{Woo2002, Sbarrato2012, Xiong2014}.

The luminosity of broad line region ($L_{\rm BLR}$) derived from broad emission lines \citep{Francis1991, Celotti1997, Sbarrato2012}, is a good estimator of the power of accretion disk, $L_{\rm Disk} \simeq 10 L_{\rm BLR}$ \citep{Calderone2013}.
Because the emission lines are produced by gas that is photoionized by the disk emission.
Thanks to \textit{Fermi}-LAT, we have come to a new era of blazar research.
\textit{Fermi} collaboration has released four generations of $\gamma$-ray source catalogues.
The fourth one (4FGL) contains 5064 sources above 4$\sigma$ signification, among these sources more than 3130 of identified or associated sources are active galaxies of blazar class (including uncertain type blazars, BCUs.) \citep{Abdollahi2020}.
Blazars are strong $\gamma$-ray emitters, their $\gamma$-ray emissions dominate bolometric luminosity ($L^{\rm bol}_{\rm jet}$) of jets. 
Thus, the $L_{\rm \gamma}$ often take the place of $L^{\rm bol}_{\rm jet}$ in previous research \citep{Ghisellini2014, Xiong2014, Zhang2020}.
The relativistic jets transport energy and momentum from AGN to large scales, but the jet formation remains unclear.
The current theoretical models consider that jet originated either from the accretion disk and powered by accretion or from the central BH and powered by extracting rotation energy \citep{Blandford1977, Blandford1982}.

The connection between relativistic jet and accretion disk through study $\gamma$-ray luminosity,  broad emission line, and blackhole mass has been explored by many authors.
\citet{Sbarrato2012} studied the blazars that have been detected by \textit{Fermi}-LAT and that are present in the Sloan Digital Sky Survey (SDSS), suggested the $L_{\rm BLR}$ correlates well with $L_{\gamma}$.
The correlation proves the emission-line photons to play a role in producing high-energy $\gamma$-rays and points out a clue of the relation between accretion ratio and jet power.
The correlations between intrinsic $\gamma$-ray luminosity and BH mass, Eddington ratio, broad-line luminosity were studied by \citet{Xiong2014} and \citet{Zhang2020}, and they all show positive correlations.
A correlation of ${\rm log}L_{\rm BLR} \sim (0.98\pm0.07){\rm log}P_{\rm jet}$ suggest that jets are powered by extraction from both accretion and BH spin \citet{Xiong2014}.

In this work, we focus on the study of the correlations between $\gamma$-ray emission and BH mass, relativistic jet related quantities to investigate the jet origination and accretion rate of blazars.
%We suggest a method by using jet power, BH mass, and disk luminosity to study the origin of jet power and propose an approach to estimate a lower limit of Doppler factor.

This paper is arranged as follows:
in Section 2, we present the samples;
the data reduction and results are presented in Section 3;
Section 4 will be our discussion;
our conclusion will be presented in Section 5.
The cosmological parameters $H_{\rm 0} = 73 \ {\rm km \cdot s^{-1} \cdot Mpc^{-1}}$, $\Omega_{\rm m} = 0.3$ and $\Omega_{\rm \Lambda} = 0.7$ have been adopted through this paper.

\section{The samples}
We collect broad emission line profiles and BH mass from \citet{Paliya2021}, which contains 674 sources.
Besides, 10 sources with emission line parameters or BH mass values from literature \citep{Baldwin1981, Chen2015} were included by \citet{Paliya2021}, and these sources are also employed in our work.
The classification of \textit{Fermi} sources are sometimes changed after a new data release. 
According to the latest classification, there were 17 sources were excluded from the blazar class, finally make us a sample of 667 sources (56 BCUs, 52 BL Lacs and 559 FSRQs). 
Meanwhile, we collect the $\gamma$-ray flux from 4FGL for the sources in our sample.
At last, we cross-correlate the sample with \citet{Nemmen2012, Ghisellini2014, Tan2020} to get the entire jet power ($P_{\rm jet}$), the non-thermal radiation power ($P_{\rm rad}$), and the accretion disk luminosity ($L_{\rm Disk}$).

\subsection{$\gamma$-ray luminosity}
The SED of blazar is characterized by two broad bumps, peaking in the mm-UV and the MeV-GeV $\gamma$-ray bands separately.
The emission of the high energy bump is usually the dominant component for blazars, so-called a higher Compton dominance that is quantified by $L_{\rm IC}/L_{\rm syn}$, except for some low power BL Lacs.
Thus, the $\gamma$-ray luminosity is believed to be a representative of blazar non-thermal bolometric luminosity \citep{Ghisellini2014, Xiong2014, Zhang2020}.
An isotropic $\gamma$-ray luminosity is expressed as
\begin{equation}
L_{\rm \gamma} = 4 \pi d_{\rm L}^2(1+z)^{\alpha_{\rm ph}-2} F,
\label{eq_L_g}
\end{equation}
where $d_{\rm L} = (1+z) \cdot \frac{c}{H_{\rm 0}} \cdot \int_{\rm 1}^{\rm 1+z} \frac{1}{\sqrt{\Omega_{\rm M}x^3+1-\Omega_{\rm M}}}dx$, $z$ is redshift, $(1+z)^{\alpha_{\rm ph}-2}$ represents a $K$-correction, $\alpha_{\rm ph}$ is the $\gamma$-ray photon index, and $F$ is the $\gamma$-ray flux in units of $\rm erg \cdot cm^{\rm -2} \cdot s^{\rm -1}$.
We calculated $L_{\rm \gamma}$ for these 637 sources, which with available redshift from the NASA/IPAC Extragalactic Database (NED), via Eq.\ref{eq_L_g}.
The redshift, 4FGL $\gamma$-ray photon density and photon spectral index that are listed in columns (3), (4) and (6) of Table \ref{tab_og}.

\subsection{BH mass and BLR luminosity}
\citet{Paliya2021} obtained emission line (H$\alpha$, H$\beta$, Mg II, and C IV) luminosity and corresponding continuum luminosity by analyzing optical spectra from SDSS-DR16 data.
Continuum luminosity $L_{\rm \lambda}$ at 5100 {\AA} is estimated from ${\rm H} \beta$ luminosity, at 3000 {\AA} is estimated from Mg II luminosity, and at 1350 {\AA} is estimated form C IV luminosity via empirical relations.
Then, a virial $M_{\rm BH}$ is estimated through empirical formula
\begin{equation}
{\rm log} M_{\rm BH} = a + b {\rm log} L_{\rm \lambda} + 2 {\rm log} {FWHM},
\label{eq_RLMass}
\end{equation}
where $M_{\rm BH}$ in units of solar mass $M_{\sun}$, $L_{\rm \lambda}$ in units of $\rm 10^{44} \ erg \cdot cm^{-2} \cdot s^{-1}$, FWHM in units of $\rm km \cdot s^{-1}$, and the calibration coefficients $a$ and $b$ are taken from \citet{McLure2004, Vestergaard2006, Shen2011}.
The BH mass are listed in column (11) of Table \ref{tab_og}.

Moreover, one can infer the luminosity of the entire broad emission line region ($L_{\rm BLR}$) from emission line luminosity.
\citet{Celotti1997} calculated $L_{\rm BLR}$ by scaling strong emission lines to the quasar template spectrum of \citet{Francis1991}.
They set Ly$\alpha$ as a reference flux that contributed to 100, the relative weight of H$\alpha$, H$\beta$, MgII and C IV lines to 77, 22, 34, and 63, and total broad line flux was fixed at 556.
The BLR luminosity is, then, expressed as
\begin{equation}
L_{\rm BLR} = \sum_{i} L_{i} \cdot \frac{\langle L_{\rm BLR, \ rel} \rangle}{\sum_{i} L_{i,\ \rm rel}},
\label{eq_L_BLR}
\end{equation}
where $\langle L_{\rm BLR,\ rel} \rangle = 556$, $L_{i}$ is observed line luminosity, and $L_{i,\ \rm rel}$ is relative line luminosity.

\subsection{Jet power}
The entire power of jet ($P_{\rm jet}$) generally contains two parts of energy, namely radiation power ($P_{\rm rad}$) and kinetic power ($P_{\rm kin}$), that in charge of its non-thermal radiation and its propagation.
%General idea is that the jet radiation $P_{\rm rad}$ would cost 10\% of its entire power $P_{\rm jet}$, and this holds for jet systems of both AGN and Gamma-Ray Burst (GRB) \citep{Nemmen2012, Ghisellini2014}.

There are methods to estimate $P_{\rm kin}$, $P_{\rm rad}$, and $P_{\rm jet}$.
\citet{Cavagnolo2010} searched for X-ray cavities in different systems including giant elliptical galaxies and cD galaxies and estimated the required jet power that is able to inflate these cavities or bubbles, obtaining a correlation between `cavity' power and radio luminosity
\begin{equation}
P_{\rm cav} \approx 5.8 \times 10^{43} \left(\frac{P_{\rm radio}}{10^{40} \ \rm erg \cdot s^{-1}} \right)^{0.7} \ \rm erg \cdot s^{-1},
\label{eq_P_cav}
\end{equation}
and assuming $P_{\rm kin} = P_{\rm cav}$. 
%This method was employed to estimate jet power in \citet{Nemmen2012} and also adopted in this work to infer $P_{\rm jet}$ from radio luminosity.
%The estimated jet power was listed in table,,,column,,,
The radiation power is expressed as
\begin{equation}
P_{\rm rad} = 2f\frac{\Gamma^{2} L_{\rm jet}^{\rm bol}}{\delta^{4}},
\label{eq_P_rad}
\end{equation}
where the factor 2 counts for two-sided jets, $f$ equals 16/5 for the case of radiation power consuming through SSC process.
For the case of EC process, $f=4/3$ and replace $\delta^{4}$ with $\delta^{4}(\delta/\Gamma)^2$.
Two assumptions, $L_{\rm bol}^{\rm jet}$ is represented by $L_{\rm \gamma}$ and $\Gamma = \delta$, are both hold for blazars \citep{Ghisellini2010, Ghisellini2014, Xiong2014, Zhang2020}.

The $P_{\rm rad}$ and $P_{\rm jet}$ are obtainable through broadband SED fitting.
Assuming that the jet power is carried by relativistic electron, cold proton, magnetic field, and radiation.
The jet power is expressed as
\begin{equation}
P_{\rm jet} = \sum_{i} \pi R^{2} {\rm \Gamma^{2}} c U_{i},
\label{eq_P_jet}
\end{equation}
where $U_{i} (i = e, p, B, rad)$ are the energy densities associated with the emitting electron $U_{\rm e}$, cold proton $U_{\rm p}$, magnetic field $U_{\rm B}$, and radiation $U_{\rm rad}$ measured in the comoving frame \citep{Ghisellini2010, Tan2020}.
We collect $P_{\rm rad}$ and $P_{\rm jet}$ from \citet{Ghisellini2014} and \citet{Tan2020} for our sources and list them in columns (4) and (5) of Table \ref{tab_jet}.

\section{Results}
\subsection{The distributions}
\begin{figure}[h]
\centering
\includegraphics[scale=0.65]{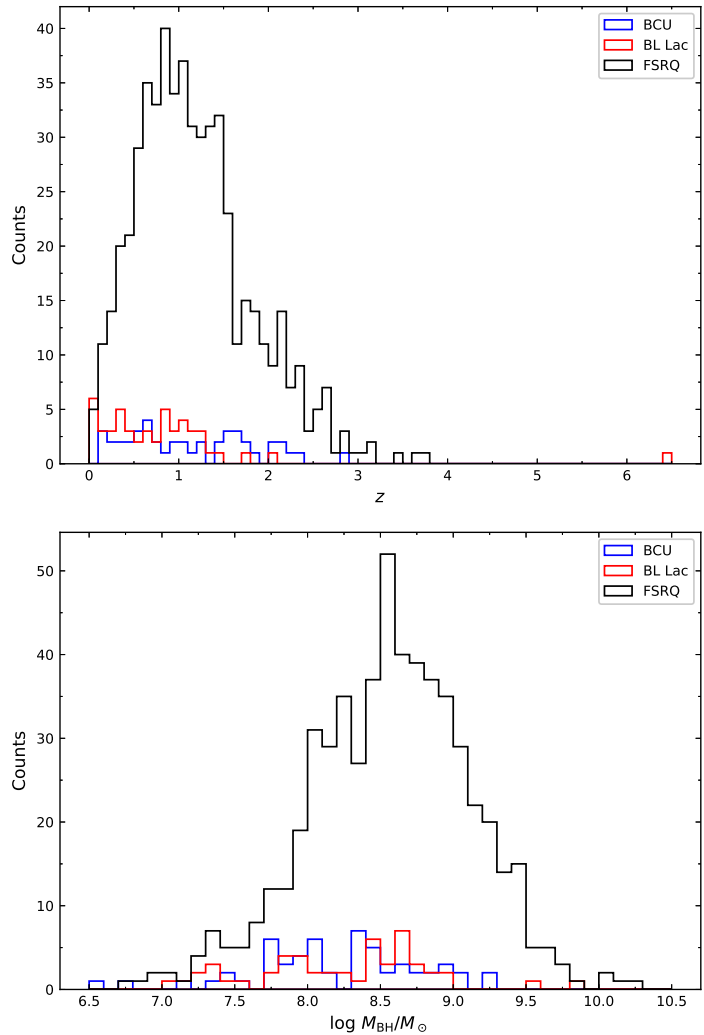}
\caption{The distributions of redshift (upper panel) and BH mass (lower panel) for the blazars of this work.
The blue histogram stands for BCU, the red one stands for BL Lac, and the black one stands for FSRQ, respectively.}
\label{Fig_hist_zM}
\end{figure}

The redshift and BH mass distributions of various classes of sources are shown in Fig. \ref{Fig_hist_zM}.
The redshift, which are obtained by checking their associate names (from 4FGL) in the NED, distributes from 0.00085 to 6.443 with a mean value of $1.147 \pm 0.688$ for all the blazars in our sample. 
The mean redshifts for FSRQs is $1.178 \pm 0.652$, for BL Lacs is $0.697 \pm 0.479$ (4FGL J0823.3+2224 is excluded for the extremely high redshift 6.443), and for BCUs is $1.144 \pm 0.698$.
The BH mass ranges from 6.35 to 10.2 with a mean value of $8.50 \pm 0.58$ for all the blazars in our sample.
The mean BH masses for FSRQs is $8.56 \pm 0.55$, for BL Lacs is $8.18 \pm 0.66$, and for BCUs is $8.19 \pm 0.57$.

\subsection{Correlation between $\gamma$-ray luminosity and BH mass}
\label{mass_formula}
\begin{figure}[h]
\centering
\includegraphics[scale=0.65]{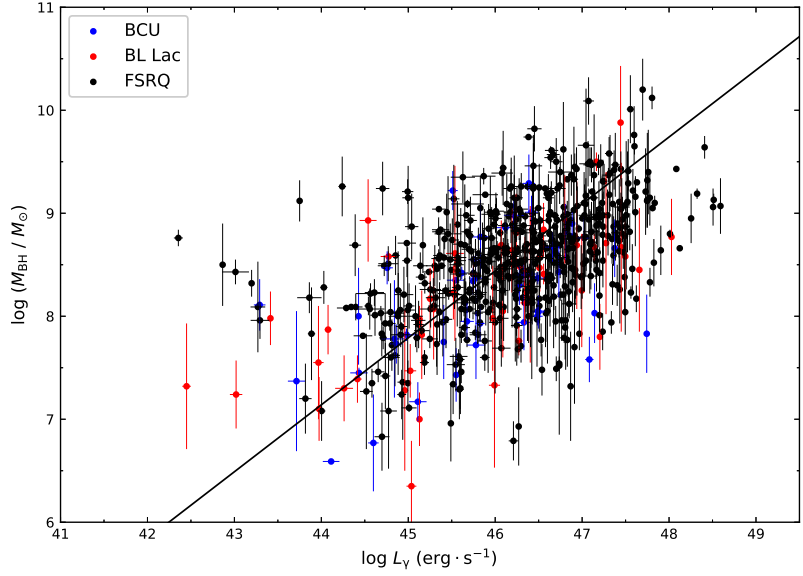}
\caption{The correlation between BH mass and $\gamma$-ray luminosity.
The black solid line stands for the result of linear regression.
The blue dot stands for BCU, red dot stands for BL Lac, and black dot stands for FSRQ, respectively.}
\label{Fig_gamMs}
\end{figure}

Fig \ref{Fig_gamMs} shows BH mass as a function of $\gamma$-ray luminosity.
We have three sources with debatable $\gamma$-ray luminosity due to their redshift, these sources are not shown in Fig \ref{Fig_gamMs} and the luminosity of these three sources will not be employed during our analysis through this paper.
Two (4FGL J1434.2+4204 and 4FGL J2134.2-0154) of them with at least two order of magnitude lower $\gamma$-ray luminosity (${\rm log}L_{\rm \gamma}=39.92$ and ${\rm log}L_{\rm \gamma}=40.38$ in unit of erg/s) than the rest of the sources due to their extreme small redshifts (0.0031 and 0.00085) with respect to the average value of their class in our sample.
In addition, we also remove the BL Lac object, 4FGL J0823.3+2224 (OJ 233), for its extremely large and suspicious redshift z=6.443.
We suspect the redshift of these three sources are mis-estimated for two possible reasons (1) optical counterparts are wrongly associated; (2) or very weak emission lines on the spectrum.
Linear regression is applied to analyse the correlation between $\gamma$-ray luminosity and BH mass for all the sources in our sample except for the above-mentioned three.
%During the analysis, the BH mass is presented in units of solar mass ($M_{\rm \sun}$).
The result indicates that BH mass and $\gamma$-ray luminosity has a strong correlation from a ordinary least squares (OLS) bisector regression
$${\rm log} \frac{M_{\rm BH}}{M_{\rm \sun}} = (0.65 \pm 0.02) {\rm log} L_{\rm \gamma} - (21.46 \pm 1.04),$$
and the correlation coefficient $r = 0.52$ and the chance probability $p = 1.3 \times 10^{-44}$ are obtained through Pearson analysis.
The result suggests a strong correlation between BH mass and $\gamma$-ray luminosity.
Thus we suggest that $\gamma$-ray luminosity is a good BH mass estimator of blazar and propose this formula
$$\frac{M_{\rm BH}}{M_{\rm \sun}} \simeq \frac{L_{\rm \gamma}^{0.65}}{21.46}.$$

\subsection{The correlation between $\gamma$-ray luminosity and BLR luminosity}
\begin{figure}[h]
\centering
\includegraphics[scale=0.65]{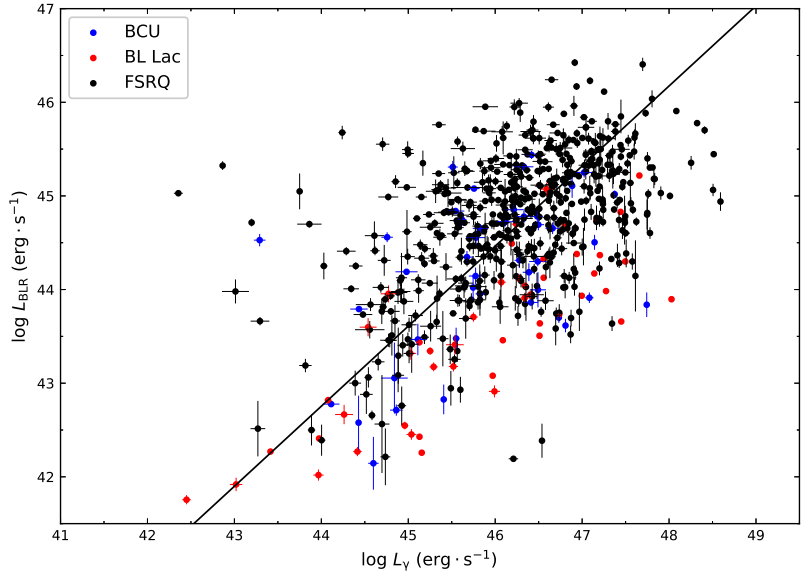}
\caption{The correlation between BLR luminosity and $\gamma$-ray luminosity.
The meaning of different symbols are as same as Fig. \ref{Fig_gamMs}}
\label{Fig_LblrLg}
\end{figure}

Fig. \ref{Fig_LblrLg} shows BLR luminosity as a function of $\gamma$-ray luminosity.
The OLS bisector regression is employed to find correlation between BLR luminosity and $\gamma$-ray luminosity, the result finds
$${\rm log} L_{\rm BLR} = (0.85 \pm 0.02) {\rm log} L_{\rm \gamma} + (5.19 \pm 1.15),$$
Pearson partial analysis indicates a $r = 0.14$ and a $p = 5.7 \times 10^{-4}$ after removing the redshift effect from these two quantities.
The result suggests that $\gamma$-ray luminosity is weakly correlated with BLR luminosity, although an apparently strong positive correlation is showing.

\subsection{The correlation between $\gamma$-ray luminosity and jet radiation power: a lower-limit of Doppler factor}
\label{delta}
\begin{figure}[h]
\centering
\includegraphics[scale=0.65]{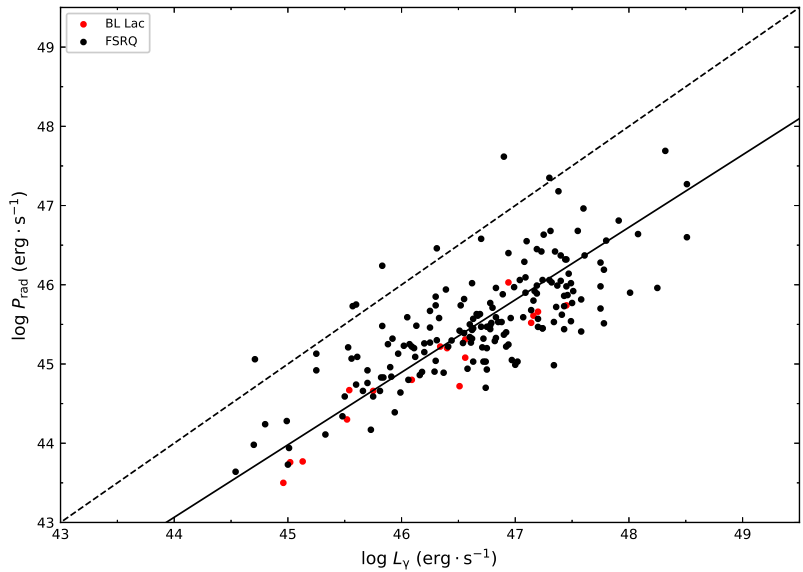}
\caption{The correlation between jet radiation power and $\gamma$-ray luminosity.
The meaning of different symbols are as same as Fig. \ref{Fig_gamMs}.
The solid black line is the linear regression, and the dashed one is the equivalent line.}
\label{Fig_LgP_1}
\end{figure}

Fig. \ref{Fig_LgP_1} shows jet radiation power as a function of $\gamma$-ray luminosity.
During the analysis, we adopt the value of $P_{\rm rad}$ from \citet{Ghisellini2014} for the common sources.
OLS bisector linear regression is employed to study the correlation between jet radiation power and $\gamma$-ray luminosity for the sources in our sample.
The regression result gives
$${\rm log} P_{\rm rad} = (0.92 \pm 0.04){\rm log} L_{\rm \gamma} + (2.81 \pm 2.05),$$
with $r=0.77$ and $p=1.6 \times 10^{-38}$, which shows the jet radiation power is strongly correlated with $\gamma$-ray luminosity.

The blazar $\gamma$-ray emission predominates its radiation power.
In fact, the $L_{\rm \gamma}$ should be less than both $P_{\rm rad}$ and $L_{\rm jet}^{\rm bol}$.
However, there are 185 sources that lie blow the equivalent line in Fig. \ref{Fig_LgP_1}, showing larger $L_{\rm \gamma}$ than $P_{\rm rad}$.
The excess of $L_{\rm \gamma}$ against $P_{\rm rad}$ suggests a significant Doppler beaming effect.

We estimate a lower-limit of Doppler beaming factor by taking two assumptions that (1) the observed $\gamma$-ray luminosity $L_{\rm \gamma}$ is a representative of $L_{\rm jet}^{\rm bol}$, and (2) $\delta$ equals to $\Gamma$ for blazars \citep{Zhang2020}.
Therefore, we have 
$$\delta = \left( 2f \frac{L_{\rm jet}^{\rm bol}}{P_{\rm rad}} \right)^{1/2} > \left( 2f \frac{L_{\rm \gamma}}{P_{\rm rad}} \right)^{1/2}.$$
The lower-limit Doppler factor ($\delta$) of our sources ranges from 3.0 to 48.6, with mean values for BL Lacs and FSRQs being $\delta_{\rm BL Lac} = 7.94 \pm 2.39$ and $\delta_{\rm FSRQ} = 11.55 \pm 6.50$, the individual values are listed in column (10) of Table \ref{tab_jet}.

\section{Discussion}
\subsection{The correlations}
BH mass is one of the key ingredients of jet origination and radiation scenarios.
There are many kinds of approaches to estimate the BH mass by using observable quantities, ie., emission line luminosity, absorption line luminosity, stellar velocity dispersion, etc \citep{Graham2007, Gultekin2009, Shen2011, Shaw2012}.
In this work, we adopt BH mass that is estimated by using emission lines to avoid variance between different methods, we collect the BH mass information from \citet{Paliya2021}.
Fig. \ref{Fig_hist_zM} show different distributions of both $M_{\rm BH}$, because we have 10 times larger sample of FSRQs than BL Lacs.
The average values indicate that FSRQs and BL Lacs have no significant difference in their BH masses, and this maybe caused by the limited number of BL Lacs in our sample.
The predicted BH masses of BL Lacs should be averagely greater than the masses of FSRQs according to the presumed `blazar cosmic evolution', in which the high-power ($L_{\rm bol} > 10^{46} \ {\rm erg \cdot s^{-1}}$) blazars (mostly FSRQs) evolve to the low-power blazars (mainly BL Lacs).
BH mass keeps growing by accretion during the evolution \citep{Cavaliere2002, Bottcher2002}, even there are different opinions about it \citep{Fan2003}. 
%The evolution suggests high-power ($L_{\rm bol} > 10^{46} \ {\rm erg \cdot s^{-1}}$) blazars, mostly FSRQs, evolve to low-power blazars, mainly BL Lacs and blazar mass keep growing by accretion \citep{Cavaliere2002, Bottcher2002}.
%Or the BH mass does not play an important role in the evolution of blazars \citep{Fan2003}.

The correlation between BH mass and $\gamma$-ray luminosity was studied by \citet{Soares2020} through using a sample of 154 FSRQs, and they proposed that the $M_{\rm BH}/M_{\rm \sun}$ is proportional to $L_{\rm \gamma}^{0.37}$.
In the present work, we have a larger sample and revisit this correlation. 
We have confirmed the positive and strong correlation between BH mass and $\gamma$-ray luminosity for blazars as shown in Fig. \ref{Fig_gamMs}.
Both \citet{Soares2020}'s and our results suggest blazar with more massive BH tend to have stronger $\gamma$-ray emission and to have a more powerful jet, and the power of jet will be discussed in section \ref{jet energy}.
However, we have a larger slope, which is 0.65, and suggest $M_{\rm BH}/M_{\rm \sun}$ proportional to $L_{\rm \gamma}^{0.65}$ that indicates BH masses may grow faster with the $\gamma$-ray luminosity than they have predicted.

The correlation between $L_{\rm \gamma}$ and $L_{\rm BLR}$ has been performed in previous studies \citep{Xiong2014, Zhang2019}.
This correlation proves that BLR provides seed photons for high energy $\gamma$-rays.
More importantly, it would point towards a relation between the accretion rate and the jet power \citep{Sbarrato2012} that we will discuss in the next section.

\subsection{A new dividing line between FSRQs and BL Lacs}
The correlation between the normalized $\gamma$-ray luminosity ($L_{\rm \gamma}/L_{\rm Edd}$, in Eddington units) and the normalized BLR luminosity ($L_{\rm BLR}/L_{\rm Edd}$) has been studied by \citet{Ghisellini2011} and \citet{Sbarrato2012}.

$L_{\rm BLR} = \xi L_{\rm Disk}$ and $L_{\rm Disk} = \eta \dot{M} c^{2}$, where $\xi$ is photoionization coefficient, $\eta$ is energy accretion efficiency, $\dot{M}$ is an accretion rate; 
$L_{\rm Edd} = \dot{M}_{\rm Edd} c^{2}$, where $\dot{M}_{\rm Edd}$ is an Eddington accretion rate.
Then we can get $\frac{L_{\rm BLR}}{L_{\rm Edd}} = \xi \eta \frac{\dot{M}}{\dot{M}_{\rm Edd}}$ by substituting $L_{\rm BLR}$ and $L_{\rm Edd}$.
If one holds $\xi$ and $\eta$ to stay constant and assumes both of them to be 0.1 as former researchers did \citep{Ghisellini2011, Sbarrato2012, Xiong2014}.
Thus, the separation is totally determined by the $\dot{M}/\dot{M}_{\rm Edd}$, which was suggested to be 0.1 refer to $L_{\rm BLR}/L_{\rm Edd} = 1 \times 10^{-3}$ \citep{Ghisellini2011}.
Later on, the value of $L_{\rm BLR}/L_{\rm Edd}$ was updated to be $5 \times 10^{-4}$ \citep{Sbarrato2012}.
The following study of \citet{Xiong2014} confirmed the idea of separation.
They concluded the boundary of accretion ratio (in Eddington units) to be $\dot{M}/\dot{M}_{\rm Edd} = 0.1$, with FSRQs showing $\dot{M}/\dot{M}_{\rm Edd} > 0.1$ and BL Lacs showing $\dot{M}/\dot{M}_{\rm Edd} < 0.1$.

\begin{figure}[h]
\centering
\includegraphics[scale=0.65]{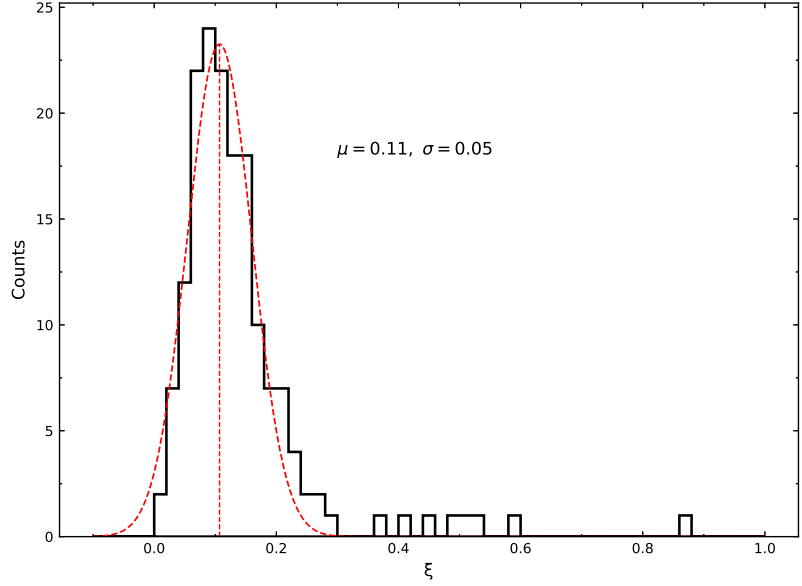}
\caption{The distribution of $\xi$ for the blazars. 
The dashed red curve stands for a Gaussian fitting of this distribution.}
\label{Fig_hist_xi}
\end{figure}

\begin{figure}[h]
\centering
\includegraphics[scale=0.65]{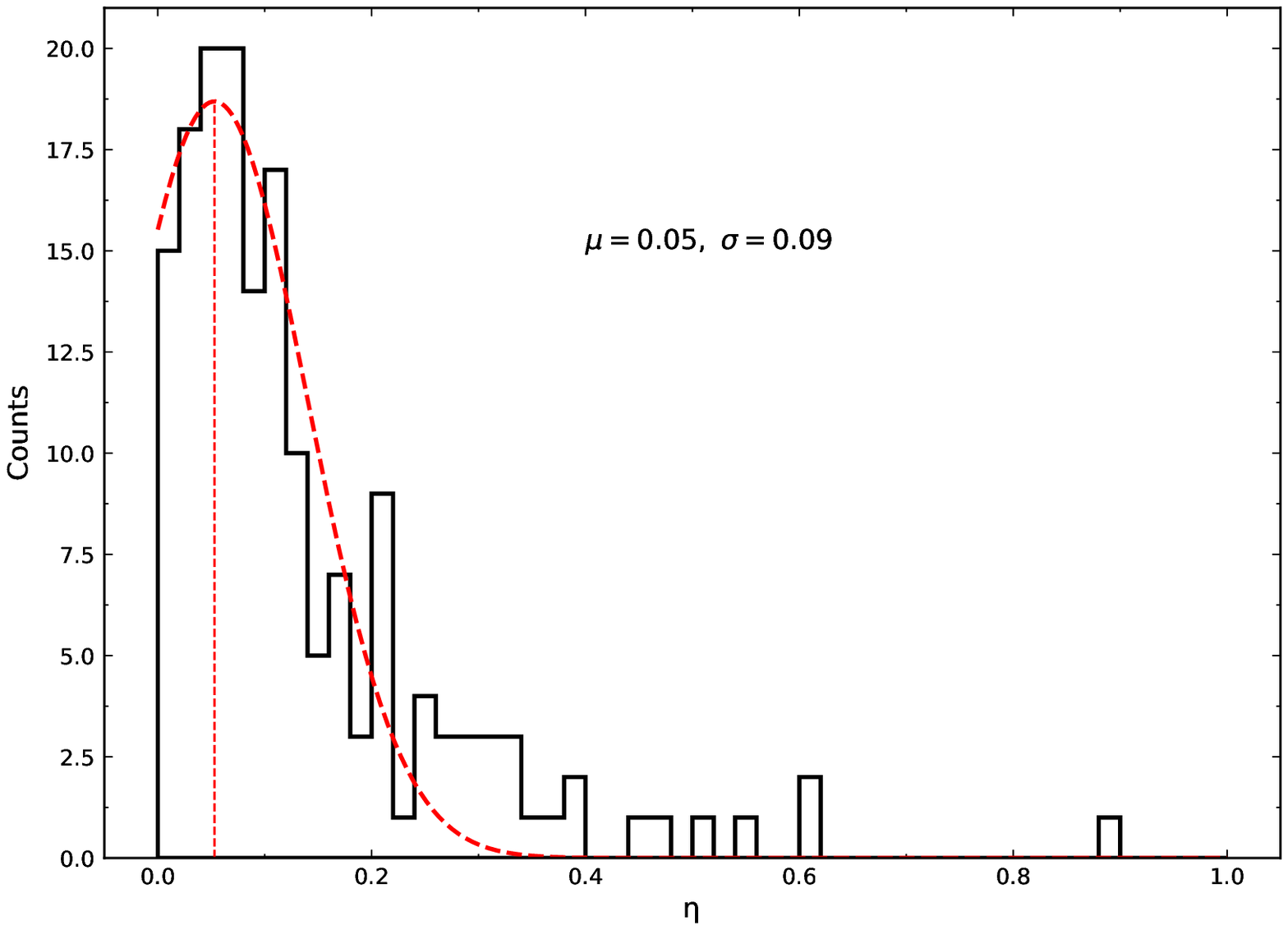}
\caption{The distribution of $\eta$ for the blazars.
The dashed red curve stands for a Gaussian fitting of this distribution.}
\label{Fig_hist_eta}
\end{figure}

While we must bear in mind that $\eta$ and $\xi$ are both assumed to be 0.1 in previous studies.
Here, we test the assumptions that $\xi = 0.1$ and $\eta = 0.1$ before we study the separation for blazars.
$\xi$ can be calculated with $L_{\rm BLR}$ and $L_{\rm Disk}$ for 166 sources (16 BL Lacs and 150 FSRQs) in our sample.
$L_{\rm BLR}$ is obtained through emission lines properties and $L_{\rm Disk}$ is obtained from \citet{Ghisellini2014}, in which they preformed SED fitting to get the disk luminosity for blazars in their sample.
The distribution of $\xi$ is shown in Fig. \ref{Fig_hist_xi}, a mean $\mu=0.11$ and standard deviation $\sigma=0.05$ are obtained when a Gaussian fitting is employed to this distribution.
The $\eta$ is difficult to estimate because it couples with $\dot{M}$, which is not able to measure directly.
A bolometric disk luminosity can be expressed as $L_{\rm Disk} = \eta \dot{M}c^{2}$, which should less than $L_{\rm Edd}$.
Thus, we can estimate a lower-limit $\eta$ because of $\eta \geq L_{\rm Disk}/L_{\rm Edd}$. 
The distribution of $\eta$ lower-limit is shown in Fig. \ref{Fig_hist_eta}, and the distribution gives a mean $\mu=0.05$ and standard deviation $\sigma=0.09$ when a Gaussian fitting is adopted to the distribution.
Our distributions of $\xi$ and $\eta$ suggest that the presumed values for both $\xi$ and $\eta$ are reasonable.

\begin{figure}[h]
\centering
\includegraphics[scale=0.65]{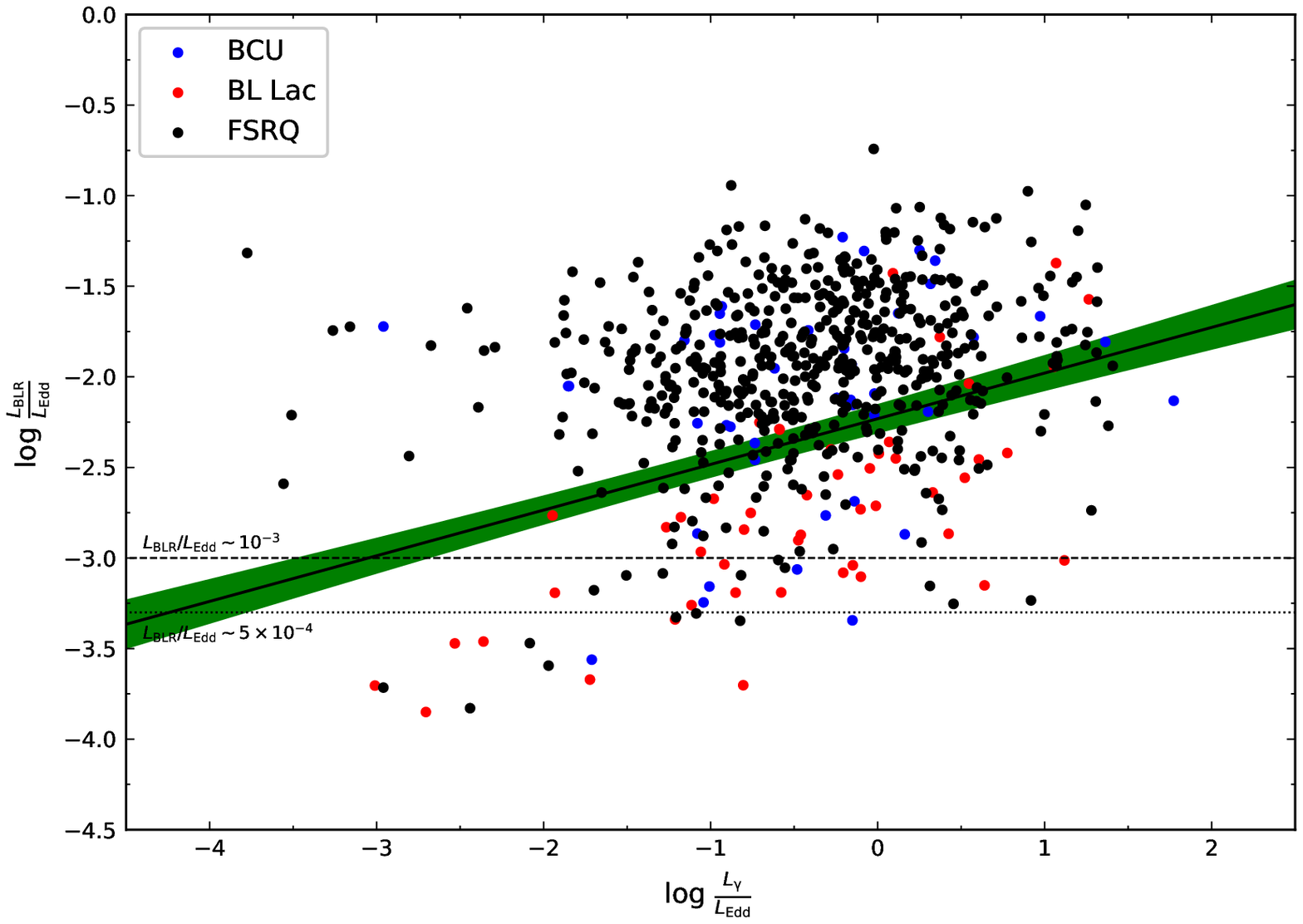}
\caption{The correlation between normalized BLR luminosity and normalized $\gamma$-ray luminosity.
The meaning of different symbols are as same as Fig\ref{Fig_gamMs}.
The solid black dividing line is our best result from SVM, the green shade plot represents its 1$\sigma$ error. 
The two horizontal lines indicate the divide between FSRQs and BL Lacs at $L_{\rm BLR}/L_{\rm Edd} \sim 10^{-3}$ from \citet{Ghisellini2011} (dashed) and at $L_{\rm BLR}/L_{\rm Edd} \sim 5 \times 10^{-4}$ from \citet{Sbarrato2012} (dotted).}
\label{Fig_div}
\end{figure}

\citet{Ghisellini2011} and \citet{Sbarrato2012} obtained dividing lines to separate FSRQs and BL Lacs.
It is interesting to revisit the dividing line using a larger sample.
We draw our sample of blazars in Fig. \ref{Fig_div} and notice many BL Lacs lying above the dividing lines that proposed by \citet{Ghisellini2011} and \citet{Sbarrato2012}.
Does that mean we need a new dividing line?
In order to do this, we employ support vector machine (SVM), a kind of machine learning (ML) method, to accomplish the task of finding a new dividing line.
The result of our dividing line gives an accuracy of 84.5\% for the separation and is expressed as
$${\rm log} \frac{L_{\rm BLR}}{L_{\rm Edd}} = 0.25 \ {\rm log} \frac{L_{\rm \gamma}}{L_{\rm Edd}} -2.23.$$

The BL Lacs lying above the dividing line have larger accretion ratio than the BL Lacs below the line, and show consequently stronger emission from BLRs.
According to the blazar evolution, we suggest these BL Lacs are at the early stage of the transition from FSRQs to BL Lacs.
On the contrary, the FSRQs below the dividing line have smaller accretion ratio are at the late stage of transition.
Moreover, we notice that there are sources, both in the above and below region, are located at the left region, ${\rm log}({L_{\rm \gamma}}/{L_{\rm Edd}}) \lesssim -2$, of the diagram.
These `left-region' sources are likely to contain a broader jet and/or a misaligned jet and show the same emission-line luminosities with respect to blazars \citep{Sbarrato2012}. 
\citet{Abdo2010c} suggested that these `left-region' sources maybe classified as radio galaxies rather than aligned blazars.
%\textbf{On the contrary, we notice that there are FSRQs lie below the dividing line in the figure.
%There are two possible explanations for the mixture of FSRQs and BL Lacs.
%Firstly, these FSRQs have smaller accretion ratio and stay at the late stage of transition according to an blazar evolution scenario;
%Secondly, for the lower-left  sources  Another possibility is that the sample used here is larger and include blazars that are less bright in gamma therefore it probably includes sources with a broader jet and/or with a jet more misaligned with respect to the line of sight.}

According to our result of the  dividing line, we believe that $\dot{M}/\dot{M}_{\rm Edd} = 0.1$ may not be a proper criteria to separate FSRQs and BL Lacs.
Instead, we suggest $\dot{M}/\dot{M}_{\rm Edd}$ evolve with the normalized $\gamma$-ray luminosity. 
%\iffalse

%\fi

\subsection{The central engine of jets}
\label{jet energy}
The origin of relativistic jets is still controversial.
\citet{Blandford1977} (BZ) used a force-free approximation and perturbative approach to study the problem of jet formation by extracting BH rotation energy.
With consideration of radiation pressure-dominated disk, one can obtain BZ-power as following
\begin{equation}
L_{\rm BZ} \simeq L_{\rm Edd} \frac{f_{+}^{2}(a)f_{\Omega}^{2}(a)}{4} \left(\frac{\beta_{\rm m}}{\alpha} \right) \left(\frac{\dot{M}c^{2}}{L_{\rm Edd}}\right)^{-3/2},
\end{equation}
where $\dot{M}$ is an accretion rate, $f_{+}(a)$ and $f_{\Omega}(a)$ are dimensionless quantities at order of 1, $\beta_{\rm m}$ is a proportion that the magnetic pressure as a fraction of the total thermodynamic disk pressure near the inner disk, and $\alpha$ gives the disk dissipation type \cite[see][Chap. 4.2]{Bottcher2012}.
Assuming that the power through the Blandford-Znajek process to be entirely transformed into jets, the sum of jet kinetic power and jet radiation power, one can express jet power as 
\begin{equation}
L_{\rm jet} \simeq L_{\rm BZ} \propto M_{\rm BH} \cdot \left(\frac{L_{\rm Disk}}{L_{\rm Edd}}\right)^{-3/2},
\label{eq_Lbz}
\end{equation}
which suggests that a slope of 1.0 for ${\rm log} L_{\rm jet}$ and ${\rm log} M_{\rm BH}$ and a slope of -3/2 for ${\rm log} L_{\rm jet}$ and ${\rm log} L_{\rm Disk}/L_{\rm Edd}$.

\begin{figure}[h]
\centering
\includegraphics[scale=0.65]{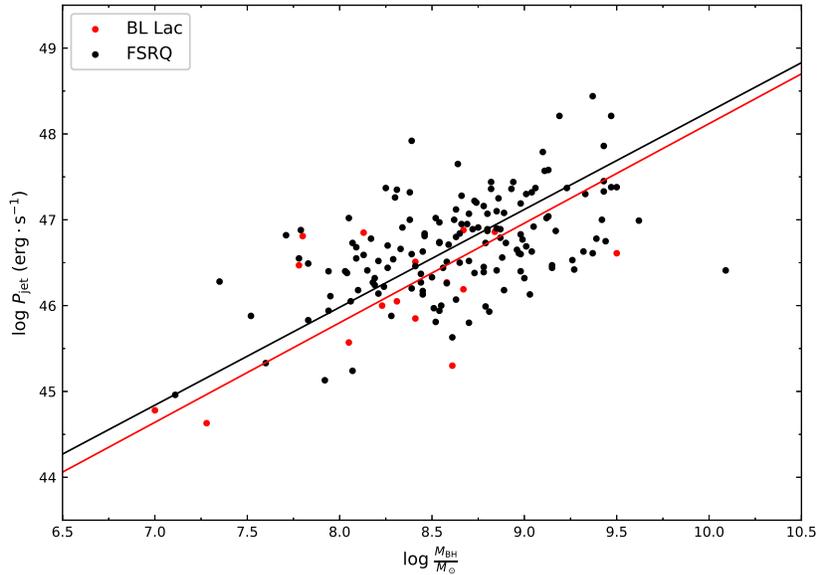}
\caption{The correlation between entire jet power and BH mass.
The meaning of different symbols are as same as Fig\ref{Fig_gamMs}.}
\label{Fig_M-Pjet}
\end{figure}

In order to investigate the nature of jet power, we collect $L_{\rm Disk}$, which estimated via modeling disk component with multi-temperature blackbody model in SED fitting procedure, from \citet{Ghisellini2014} and $P_{\rm jet}$, which estimated via SED fitting, from \citet{Ghisellini2014} and \citet{Tan2020}.
When linear regressions are used for the correlation between the jet power and BH mass, significant correlations are obtained and shown in Fig. \ref{Fig_M-Pjet}
$${\rm log} P_{\rm jet} = (1.16 \pm 0.15) {\rm log} \frac{M_{\rm BH}}{M_{\sun}} + (36.52 \pm 1.24),$$
and $r=0.59$ and $p=0.02$ for BL Lacs; and 
$${\rm log} P_{\rm jet} = (1.14 \pm 0.07) {\rm log} \frac{M_{\rm BH}}{M_{\sun}} + (36.86 \pm 0.62),$$
and $r=0.51$ and $p=1.9 \times 10^{-11}$ for FSRQs.
The results demonstrate strong correlations between the two quantities and suggest positive correlations between $M_{\rm BH}$ and $L_{\rm \gamma}$ for both BL Lacs and FSRQs. 
Moreover, slopes of $1.16\pm0.15$ and $1.14\pm0.07$ are consistent with the theoretically predicted slope that is 1.0 following Equation \ref{eq_Lbz}.

\begin{figure}[h]
\centering
\includegraphics[scale=0.65]{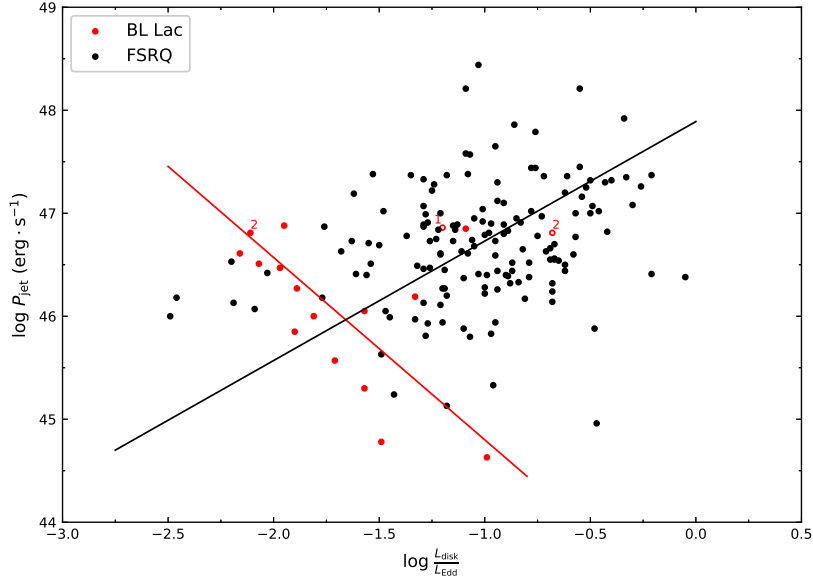}
\caption{The correlation between entire jet power and disk luminosity divided by Eddington luminosity.
The meaning of different symbols are as same as Fig\ref{Fig_gamMs}.
The solid lines are represented for the linear regressions for FSRQs and BL Lacs.
4FGL J0407.5+0741 and 4FGL J0438.9-4521 are marked as `1' and `2'.}
\label{Fig_Pr}
\end{figure}

Fig. \ref{Fig_Pr} shows the diagram of entire jet power $P_{\rm jet}$ against normalized disk luminosity $L_{\rm Disk}/L_{\rm Edd}$.
It is found that there are 2 BL Lacs in red open circles, 4FGL J0407.5+0741 and 4FGL J0438.9-4521, that are marked as `1' and `2' respectively.
4FGL J407.5+0741, known as TXS 0404+075, is a BL Lac class $\gamma$-ray emission object.
However, this source is classified as FSRQs in other studies. 
\citet{Tan2020} suggested an external Compton model, which is usually applied to FSRQs due the existence of a BLR or a dust torus, to fit its broad band SED and studied the physical properties of FSRQs. 
\citet{Xiong2014} classified this source as a LSP, which is mostly consist of FSRQs, during their study of intrinsic $\gamma$-ray luminosity and jet power.
This source shows typical a broad SED of FSRQ type, meanwhile, it shows the BL Lac optical spectrum.
Therefore, the exact classification of this source is in debate, it is better to exclude this source when investigate the possible physical property difference between FSRQs and BL Lacs.
4FGL J0438.9-4521 has a redshift of 2.017 and a black hole mass of ${\rm log}(M_{\rm BH}/M_{\sun})=7.8$.
This black hole mass is relatively small with respect to the average values of the three classes in our sample.
We notice that its BH mass was obtained according to the $C_{\rm IV} (\lambda =1549 \ {\rm \AA})$ emission line profile.
However, the infrared emission could be significantly absorbed by the dust from the local to the galaxy itself, especially, the absorption from the latter one can hardly be measured and compensated.
We believe that the mass of this source could be underestimated due to its lower emission line luminosity of $C_{\rm IV}$.
We re-calculate the $M_{\rm BH}$ via the method that we have proposed in section \ref{mass_formula} for 4FGL J0438.9-4521 and obtain a ${\rm log}(M_{\rm BH}/M_{\sun})=9.23$.
Then the plot is updated with a red dot marked as `2' with an updated BH mass. 
Linear regressions are applied independently for BL Lacs and FSRQs
$${\rm log} P_{\rm jet} = -(1.77 \pm 0.40) {\rm log} \frac{L_{\rm Disk}}{L_{\rm Edd}} + (43.03 \pm 0.68),$$
with $r=-0.52$ and $p=0.04$ for BL Lacs;
%This source has both high jet power and accretion ratio, we suggest this source at the early stage of blazar evolution.
%Indeed, this source is lying within our dividing line 1 $\sigma$ error (green shade) area in Fig. \ref{Fig_div}.
$${\rm log} P_{\rm jet} = (1.16 \pm 0.06) {\rm log} \frac{L_{\rm Disk}}{L_{\rm Edd}} + (47.89 \pm 0.62),$$
with $r=0.27$ and $p=8.5 \times 10^{-4}$ for FSRQs.
The result of slope for BL Lac, $-1.77\pm0.40$, reaches the expected slope -3/2 following Equation \ref{eq_Lbz}, indicates jets are powered by extracting BH rotation energy for BL Lacs.
A positive correlation with slope 1.16 for FSRQs suggests jets power comes from at least a mixture of extracting BH rotation power and disk accretion power, and the disk accretion power may be the dominant one.
%The result is consistent with the scenario that FSRQs contain more powerful accretion disks than BL Lacs.
%In this case, FSRQs feed jet through extracting accretion energy via magnetohydrodynamic (MHD) process is promising.

The energy extraction from BH was well established by \citet{Blandford1977}, and the following studies suggest that this process works in blazars and radio-loud narrow line Seyfert 1 AGNs (NLS1s) \citep{Xiong2014, Foschini2014}.
\citet{Xiong2014} studied the subject of blazar jet power through an investigation on the correlation between ${\rm log}L_{\rm BLR}$ and ${\rm log}P_{\rm jet}$ and obtained a slope $0.98 \pm 0.07$ for this correlation.
Their results was perfectly consistent with the theoretically predicted slope 1 for  ${\rm log}L_{\rm BLR}$ vs ${\rm log}P_{\rm jet}$, and suggested that \textit{Fermi} blazars jets powered through the BZ mechanism.

In the present work, we have confirmed that the BZ mechanism makes great efforts in \textit{Fermi} blazar jets powering. 
Moreover, our results seems to suggest BL Lacs maybe powered mostly by the BZ process that extracting energy from BH rotation.
And the BL Lac jets are likely governed by the BH spin.
While this result should be carefully used because we only have a small sample of 16 BL Lacs to study the correlation between these two quantities.
For FSRQs, our results suggest that their jets are powered mostly by the accretion disk.
And FSRQs jets raise from the inner region of accretion and the energy transformed through the magnetic field.

\subsection{Doppler beaming effect}
Blazars are known to show extreme observation properties that are associated with the Doppler beaming effect.
The beaming effect arises from the preferential orientation of the jet, typically within $< 20^{\circ}$ from our line of sight \citep{Readhead1978, Blandford1979, Readhead1980}.
This effect is quantified by a Doppler factor ($\delta$), $\delta = [\Gamma(1-\beta cos \theta)]^{-1}$, where $\Gamma$ is bulk Lorentz factor ($\Gamma = 1/\sqrt{1-\beta^{2}}$), $\beta$ is velocity of the jet in units of speed of light, and $\theta$ is the viewing angle.
Since there is no direct way to measure $\beta$ or $\theta$, $\delta$ can only be estimated by indirect methods \citep{Hovatta2009, Fan2013, Fan2014, Ghisellini2014, Chen2018, Liodakis2018, Zhang2020}.
While different methods often yield discrepant results.
\citet{Hovatta2009} and \citet{Liodakis2018} determined the variability Doppler factor at radio band by analyzing baazar light curves.
However, the estimation at radio bands may not suitable for using in $\gamma$-ray bands.
Because the $\gamma$-ray emission is extremely variable and $\gamma$-ray emission have different mechanisms, which are synchrotron-self Compton (SSC) process and external Compton (EC) process in the leptonic scenario, compare to the radio emission mechanism, which is the synchrotron process.
Doppler factor estimation at $\gamma$-ray band was proposed by \citet{Zhang2020}, $\delta^{\rm Z20}$ can be calculated through $L_{\rm \gamma}$ and $L_{\rm BLR}$ for FSRQs and BL Lacs respectively, ${\rm log} \delta^{\rm Z20}_{\rm FSRQ} = ({\rm log} L_{\rm \gamma} - 1.18 {\rm log} L_{\rm BLR} + 8.00)^{0.5}$ and ${\rm log} \delta^{\rm Z20}_{\rm BL Lac} = ({\rm log} L_{\rm \gamma} + 0.87 {\rm log} L_{\rm BLR} + 6.23)^{0.5}$.

\begin{figure}[h]
\centering
\includegraphics[scale=0.65]{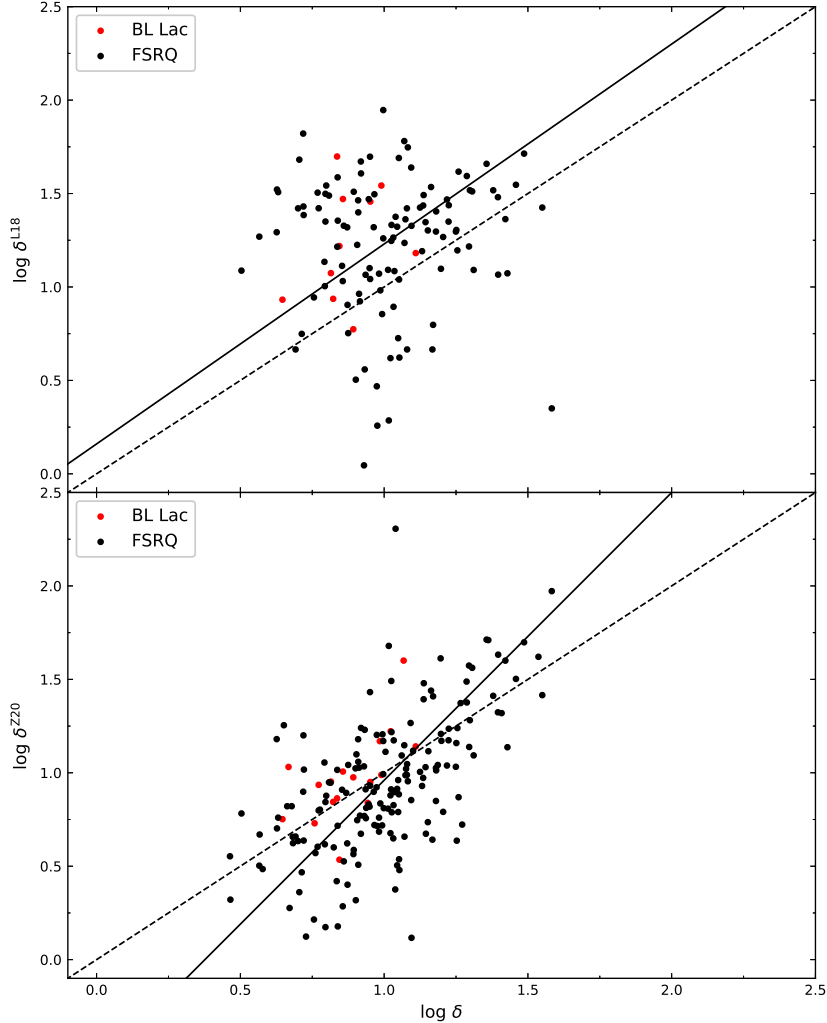}
\caption{The comparison between the Doppler factor calculated in this work (${\rm log} \delta$) and Doppler factors from \citet{Liodakis2018} (${\rm log} \delta^{\rm L18}$) and from \citet{Zhang2020} (${\rm log} \delta^{\rm Z20}$).
The meaning of different symbols are as same as Fig. \ref{Fig_gamMs}.
The dashed lines are the corresponding equivalent lines.}
\label{Fig_delta}
\end{figure}

In order to make comparisons with $\delta$ that we have calculated in section \ref{delta}.
We calculate $\gamma$-ray Doppler factor using \citet{Zhang2020}'s method for the sources in our sample, see as $\delta^{\rm Z20}$ in Tab. \ref{tab_jet} column (9).
A Kolmogorov–Smirnov (K-S) test is applied to test if $\delta$ and $\delta^{\rm Z20}$ are from the same distribution.
The K-S test result gives $p=1.8 \times 10^{-14}$ that indicates $\delta$ and $\delta^{\rm Z20}$ are from two different distributions, and implies that the method we estimate $\delta$ is independent from \citet{Zhang2020}'s to estimate $\delta^{\rm Z20}$.
$\delta^{\rm Z20}$ ranges from in 1.31 to 202.31 with mean values of $\langle \delta^{\rm Z20}_{\rm FSRQ} = 13.36 \rangle$ and $\langle \delta^{\rm Z20}_{\rm BL Lac} = 11.02 \rangle$.
We suggest that $\delta$ and $\delta^{\rm Z20}$ are comparable for two reasons (1) the average value of $\delta^{\rm Z20}$ within the one $\sigma$ error of $\delta$, (2) the data points for the common sources are almost equally distributed below and above the equivalent line in the lower panel of Fig. \ref{Fig_delta}.
This result is expected because these two kinds of Doppler factors are both obtained at $\gamma$-ray band and both use $\gamma$-ray luminosity.
%Because we have assumed $P_{\rm rad}$ equals to $L_{\rm in}$, however, $P_{\rm rad}$ actually supply to emission from radio to $\gamma$-ray band.

Besides, we collect Doppler factor from \citet{Liodakis2018}, see as $\delta^{\rm L18}$ in Tab. \ref{tab_jet} column (8).
Similarly, a K-S test with $p=6.3 \times 10^{-10}$ suggests $\delta$ and $\delta_{\rm L18}$ are from two different distributions, and implies that our method to estimate $\delta$ is independent from \citet{Liodakis2018}'s.
$\delta^{\rm L18}$ ranges from 1.11 to 88.44 with $\langle \delta^{\rm L18}_{\rm FSRQ} = 22.46 \rangle$ and $\langle \delta^{\rm L18}_{\rm BL Lac} = 20.97 \rangle$.

Among $\delta$, $\delta^{\rm L18}$ and $\delta^{\rm Z20}$, $\delta^{\rm L18}$ has the largest average values for both FSRQs and BL Lacs.
$\delta^{\rm L18}$ was driven from short term radio variability, while $\delta^{\rm Z20}$ and $\delta$ are calculated through using 8-year $\gamma$-ray average flux, in which the rapid variability information had been washed out.
One should keep in mind that a higher variability and shorter variability timescale yield a larger Doppler factor.
Consequently, $\delta^{\rm L18}$ has the largest average value among $\delta^{\rm L18}$, $\delta^{\rm Z10}$, and $\delta$ in this work.

Correlations between $\delta$ and $\delta^{\rm Z20}$, and $\delta^{\rm L18}$ for 122 common sources are shown in Fig. \ref{Fig_delta}.
A correlation between ${\rm log} \delta$ and ${\rm log} \delta^{\rm Z20}$ suggests our lower-limits of $\gamma$-ray Doppler factor is consistent with \citet{Zhang2020}'s, because we both estimate Doppler factors in the $\gamma$-ray band.
However, our result is barely correlated with \citet{Liodakis2018}'s result due to different methods and wavelength bands that have been employed.

%\iffalse
%\fi

\section{Conclusion}
In order to study blazar jet properties and its central engine, in this work, we have obtained a sample of 667 \textit{Fermi} blazars that with available emission lines profiles, $\gamma$-ray emission and SED information from the literature.
We have studied the correlations between $\gamma$-ray luminosity and BH mass, BLR luminosity and jet power, then further discussed accretion ratio separation of blazars, the jet origination, and proposed a new method of a lower-limit Doppler factor estimation.

Our main results are following:
(1) The analysis between BH mass and the $\gamma$-ray luminosity show a strong correlation in logarithmic space.
We propose a method to estimate the BH mass from $\gamma$-ray luminosity that expressed as $M_{\rm BH}/M_{\rm \sun} \simeq L_{\rm \gamma}^{0.65}/21.46.$
(2) The correlation between BLR luminosity and $\gamma$-ray luminosity is weak.
Then we normalize these two quantities with Eddington luminosity, and generate a dividing line to separate FSRQs and BL Lacs via the ML method.
The dividing line is a symbol of the accretion ratio, we suggest the accretion ratio is evolved with normalized $\gamma$-ray luminosity.
(3) Through the study between jet power and BH mass, and disk luminosity.
We have confirmed that the BZ mechanism works in the BH-disk system of blazars. Specifically, the BL Lacs jets are likely powered mainly from extracting BH rotation energy while FSRQs jets are mostly powered by an accretion disk.
(4) We propose a method of estimating a lower-limit of Doppler factor using $L_{\rm \gamma}$ and $P_{\rm rad}$, give average values $\delta_{\rm FSRQ} = 11.55 \pm 6.50$ and $\delta_{\rm BL Lac} = 7.94 \pm 2.39$.

\begin{acknowledgments}
We thank the support from our laboratory, the key laboratory for astrophysics of Shanghai, and we thank Dr. Vaidehi S. Paliya for sharing data and exchanging ideas of this work.
Meanwhile, L. P, Fu acknowledges the support from the National Natural Science Foundation of China (NSFC) grants 11933002, STCSM grants 18590780100, 19590780100, SMEC Innovation Program 2019-01-07-00-02-E00032 and Shuguang Program 19SG41.
S. H, Zhang acknowledges the support from by Natural Science Foundation of Shanghai (20ZR1473600).
J. H, Fan acknowledges the support by the NSFC (NSFC 11733001, NSFC U2031201, NSFC U1531245).

\end{acknowledgments}

\begin{table}[htbp]\scriptsize
\centering
\caption{Optical and $\gamma$-ray parameters}
\label{tab_og}
\begin{tabular}{lcccccccccc}
\hline
4FGL name & Class & z & ${\rm F}_{\gamma}$ & ${\rm Unc}\_ {\rm F}_{\gamma}$ & $\Gamma_{\rm ph}$ & ${\rm log} L_{\rm H \alpha}$ & ${\rm log} L_{\rm H \beta}$ & ${\rm log} L_{\rm Mg II}$ & ${\rm log} L_{\rm C IV}$ & ${\rm log} M_{\rm BH}/M_{\sun}$ \\
(1) & (2) & (3) & (4) & (5) & (6) & (7) & (8) & (9) & (10) & (11)  \\
\hline
J0001.5+2113	&	F	&	1.106	&	1.36E-09	&	6.86E-11	&	2.66	&	$42.942 \pm 0.016$	&	$42.029 \pm 0.054$	&	$42.503 \pm 0.032$	&		&	$7.54 \pm 0.07$	\\
J0004.3+4614	&	F	&	1.81	&	2.41E-10	&	3.92E-11	&	2.58	&		&		&		&	$44.126 \pm 0.031$	&	$8.36 \pm 0.1$	\\
J0004.4-4737	&	F	&	0.88	&	4.36E-10	&	3.75E-11	&	2.37	&		&		&	$42.885 \pm 0.099$	&		&	$8.28 \pm 0.27$	\\
J0006.3-0620	&	B	&	0.346676	&	1.40E-10	&	3.13E-11	&	2.13	&	$42.782 \pm 0.189$	&	$42.004 \pm 0.227$	&		&		&	$8.93 \pm 0.4$	\\
J0010.6+2043	&	F	&	0.5978	&	1.73E-10	&	3.44E-11	&	2.32	&		&	$43.047 \pm 0.048$	&	$43.027 \pm 0.017$	&		&	$7.86 \pm 0.04$ \\
\hline
\end{tabular}
\tablecomments{Column definitions: 
(1) 4FGL name;
(2) Classification, `B' stands for BL Lacs, `F' stands for FSRQs, `U' stands for BCUs;
(3) redshift;
(4) integral photon flux from 1 to 100 GeV, in units of ${\rm photon \cdot cm^{-2} \cdot s^{-1}}$; 
(5) 1 $\sigma$ error of $F_{\rm \gamma}$;
(6) photon index; 
(7) luminosity of H$\alpha$ emission line, in units of ${\rm erg \cdot cm^{-1} \cdot s^{-1}}$; 
(8) luminosity of H$\beta$ emission line in units of ${\rm erg \cdot cm^{-1} \cdot s^{-1}}$; 
(9) luminosity of Mg II emission line in units of ${\rm erg \cdot cm^{-1} \cdot s^{-1}}$; 
(10) luminosity of C IV emission line in units of ${\rm erg \cdot cm^{-1} \cdot s^{-1}}$; 
(11) BH mass, in units of solar mass.
Only 5 objects are presented here, the table is available in its entirety in machine-readable form.}
\end{table}

\begin{table}[htbp]\scriptsize
\centering
\caption{Jet parameters}
\label{tab_jet}
\begin{tabular}{lccccccccc}
\hline
4FGL name & Class & z & ${\rm log} P_{\rm rad}$ & ${\rm log} P_{\rm jet}$ & ${\rm log} L_{\rm Disk} \ (\rm SED)$ & Ref. & $\delta^{\rm L18}$ & $\delta^{\rm Z20}$ & $\delta$ \\
(1) & (2) & (3) & (4) & (5) & (6) & (7) & (8) & (9) & (10)  \\
\hline
J0001.5+2113	&	F	&	1.106	    &		    &		    &		    &		&		    &	40.77   &	    	\\
J0004.3+4614	&	F	&	1.81	    &		    &		    &		    &		&	7.75	&	5.63    &		    \\
J0004.4-4737	&	F	&	0.88	    &	44.64 	&	45.88 	&	45.32	&	G14	&		    &	10.50 	&	11.92 	\\
J0006.3-0620	&	B	&	0.346676	&		    &		    &		    &		&	6.96	&	2.48    &	    	\\
J0010.6+2043	&	F	&	0.5978	    &		    &		    &		    &		&	6.02	&	2.92    &	    	\\
\hline
\end{tabular}
\tablecomments{Column definitions: 
(1) 4FGL name; 
(2) Classification, `B' stands for BL Lacs, `F' stands for FSRQs, `U' stands for BCUs; 
(3) redshift; 
(4) jet radiation power, in units of ${\rm photon \cdot cm^{-2} \cdot s^{-1}}$; 
(5) jet entire power, in units of ${\rm photon \cdot cm^{-2} \cdot s^{-1}}$;
(6) luminosity of accretion disk, in units of ${\rm erg \cdot cm^{-1} \cdot s^{-1}}$;
(7) references of $P_{\rm rad}$, $P_{\rm jet}$, and $L_{\rm Disk}$, that `G14' sands for \citet{Ghisellini2014} and `T20' stands for \citet{Tan2020};
(8) Doppler factor from \citet{Liodakis2018};
(9) estimated Doppler factor using the method proposed in \citet{Zhang2020};
(10) estimated lower-limit Doppler factor in the present work.
Only 5 objects are presented here, the table is available in its entirety in machine-readable form.}
\end{table}

\bibliography{lib}{}
\bibliographystyle{aasjournal}



\end{document}